\documentclass[aps,prl,groupedaddress]{revtex4}
\usepackage{amsmath}
\usepackage{epsfig,epsf}
\usepackage{graphicx}
\usepackage{verbatim}
\usepackage{epstopdf}

\def\beq{\begin{equation}}
\def\eeq{\end{equation}}

\begin{document}


\voffset1.5cm
\title{ Jalilian-Marian, Iancu, McLerran, Weigert, Leonidov, Kovner evolution at next to leading order.}
\author{ Alex Kovner$^1$, Michael Lublinsky$^2$ and Yair Mulian$^2$}
\affiliation{
$^1$Physics Department, University of Connecticut, 2152 Hillside road, Storrs, CT 06269, USA\\
$^2$ Physics Department, Ben-Gurion University of the Negev, Beer Sheva 84105, Israel\\}
 
\date{\today}
\begin{abstract}
The  Jalilian-Marian,Iancu, McLerran, Weigert, Leonidov, Kovner (JIMWLK) Hamiltonian for high energy evolution of QCD amplitudes is presented at the next-to-leading order accuracy in $\alpha_s$. The form of the Hamiltonian is deduced from the symmetries and  the structure of the hadronic light cone wavefunction and by comparing the rapidity evolution of the quark dipole and the three-quark singlet states  with results  available in the literature. The next-to-leading corrections should allow for more robust phenomenological applications of perturbative saturation approach. 
 \end{abstract}
\maketitle

It is believed that at high energy gluons saturate in perturbative regime. The idea of perturbative gluon saturation was first suggested and discussed in detail in \cite{GLR}. To date there exist numerous phenomenological applications of this idea to DIS, heavy ion 
collisions and proton-proton collisions at the LHC \cite{review}. These applications are based on the Balitsky-Kovchegov  (BK) non-linear evolution equation \cite{Bal,KOV}, which at large $N_c$ describes the growth of the gluon density with energy and the gluon saturation. The more general approach to the calculation of high energy hadronic amplitudes is known as the Jalilian-Marian,Iancu, McLerran, Weigert, Leonidov, Kovner (JIMWLK) evolution.
The JIMWLK Hamiltonian \cite{jimwlk} is the limit of the QCD Reggeon Field Theory (RFT),
applicable for computations of high energy  scattering amplitudes   of dilute (small parton number) projectiles on dense (nuclei) targets.  In general it predicts the rapidity evolution of any hadronic observable $O$ via the functional equation of the form
\beq\label{1}
\frac{d}{dY}\,{\cal O} \,=\,-\,H^{JIMWLK}\,{\cal O}
\eeq
In ref. \cite{jimwlk}, the JIMWLK Hamiltonian was derived in the leading order in $\alpha_s$ in pQCD. It  contains a wealth of information about high energy evolution equations.
In the dilute-dilute limit it  generates the linear  BFKL equation \cite{bfkl} and
its BKP extension \cite{bkp}. Beyond the dilute limit, the Hamiltonian incorporates non-linear effects responsible for unitarization of scattering amplitudes. 

The BK equation arises as the mean field approximation to JIMWLK evolution at large $N_c$.
Successful BK phenomenology mandates inclusion of next to leading order corrections, since at leading order the evolution predicted by the BK equation is too rapid to describe experimental data. Currently only the running coupling corrections are included in applications, although it is clearly desirable to include all next to leading corrections. The complete set of such corrections to the evolution of a fundamental dipole was calculated in a remarkable paper by Balitsky and Chirilli \cite{BC}. The complete functional JIMWLK equation however, at the moment is  only known at leading order. The next to leading order extension of the JIMWLK framework is imperative for calculation of more general amplitudes, beyond the dipole, which determine interesting experimental observables like single- and double inclusive particle production\cite{footnote}. 

Beyond phenomenological interest, deriving and exploring RFT at NLO is a fundamental theoretical question and it is the focus of this paper. 

The NLO  BFKL equation was derived in \cite{NLOBFKL}. Its perturbative solution was recently presented in \cite{CK}.  The odderon and the linear BKP hierarchy at NLO were studied recently in \cite{NLOBKP}.
The extension of the NLO BFKL which fully resums multiple scattering unitarizing effects was achieved in \cite{BC}, following on the earlier works \cite{BalNLO}.  Ref.\cite{BC} put forward the evolution equation of a quark color dipole operator at NLO.  The subset of these corrections - the running coupling effects in the BK equation were also calculated in \cite{Weigertrun}. The photon impact factor at NLO was computed in \cite{BCimpact}.
Recently Grabovsky \cite{Grab} computed certain  parts of the NLO evolution equation for three-quark 
singlet amplitude in the $SU(3)$ theory.  Despite this major progress in the NLO computations, the complete NLO JIMWLK Hamiltonian, which must generate these results by simple application to the relevant amplitudes, has not been constructed yet.  
Our goal here is to complete this step: we construct the JIMWLK Hamiltonian which unifies  the previously available results and is the starting point to generalization thereof. 

Our approach is to use two major pieces of input. First, the general form
of the NLO JIMWLK Hamiltonian can be deduced from the hadronic  wavefunction computed in the light cone perturbation
theory \cite{LM}. This results in parametrization of the Hamiltonian in terms of five kernels. Second, these kernels are fully reconstructed by comparing
the evolution generated by the Hamiltonian with the detailed results of \cite{BC} 
and \cite{Grab}. Via this route we determine the kernels, deduce the complete Hamiltonian and with its help perform some non-trivial consistency crosschecks between the 
results of \cite{BC} and \cite{Grab}. 
 
In this paper we  outline the basic steps in our derivation and present the final form of the Hamiltonian. 
Major elements of the derivation are presented in the Appendix. Additional details and expanded discussion 
can be  found in our recent publication \cite{KLMconf}. 
In the upcoming publication \cite{complete} we will provide a detailed comparison of our approach with \cite{BClast} which appeared simultaneously with the first version of the present paper.

The JIMWLK Hamiltonian defines a two-dimensional non-local field theory of a  unitary matrix (Wilson line) $S(x)$ which, in the high energy eikonal approximation represents the scattering amplitude of a quark at the transverse coordinate $x$.  
The leading order  Hamiltonian is:
\begin{eqnarray}\label{LO}
H^{LO\ JIMWLK}\ = \ \int d^2 z\,d^2 x\,d^2 y\, K^{LO}(x,y,z)\ \left[ J^a_L(x)\,J^a_L(y)\,+\,J_R^a(x)\,J_R^a(y)\,-\,2\,J_L^a(x)\,S^{ab}_A(z)\,J^b_R(y)\right]
\end{eqnarray}
The left and right $SU(N)$ rotation generators, when acting on functions of $S$ have the representation
\begin{eqnarray}\label{LR}
J^a_L(x)=tr\left[\frac{\delta}{\delta S^{T}_x}T^aS_x\right]-tr\left[\frac{\delta}{\delta S^{*}_x}S^\dagger_xT^a\right]  \,; \ \ \ \ \ \ \ \ 
J^a_R(x)=tr\left[\frac{\delta}{\delta S^{T}_x}S_xT^a\right] -tr\left[\frac{\delta}{\delta S^{*}_x}T^aS^\dagger_x\right]
\end{eqnarray}
Here $T^a$ are $SU(N)$ generators in the fundamental representation, while $S_A$ is a unitary matrix in the adjoint representation - the gluon scattering amplitude. The leading order kernel  is given by
\beq
K^{LO}(x,y,z)\,=\,\,\frac{\alpha_s}{2\,\pi^2}\,\frac{X\cdot Y}{ X^2\,Y^2}
\eeq
We use the notations of ref. \cite{BC} $X\equiv x-z$,  $X^\prime\equiv x-z^\prime$, $Y\equiv y-z$,    $ Y^\prime \equiv y-z^\prime$,
$W\equiv w-z$,  and  $ W^\prime \equiv w-z^\prime$.

The LO Hamiltonian is invariant under $SU_L(N)\times SU_R(N)$ rotations, which reflects gauge invariance of scattering amplitudes.
When acting on gauge invariant operators (operators invariant separately under $SU_L(N)$ and $SU_R(N)$ rotations), the kernel $K^{LO}$ can be 
substituted by the so called dipole kernel \cite{Iancu}
\begin{equation}
K^{LO}(x,y,z)\ \rightarrow \ - \frac{1}{2}M(x,y;z); \ \ \ \ \ \ \ \ M(x,y;z)\,=\,\frac{\alpha_s}{2\,\pi^2}\,\frac{(x-y)^2}{X^2\,Y^2}
\end{equation}
which vanishes at $x=y$ and has a better IR behavior. In addition, the Hamiltonian is invariant under the $Z_2$ transformation $S\rightarrow S^\dagger; \ \ J_L\rightarrow -J_R$, which in \cite{reggeon} was identified as signature, and the charge conjugation symmetry $S\rightarrow S^*$.

The JIMWLK Hamiltonian is derivable from perturbatively computable hadronic wavefunction \cite{KL}. At LO, the wavefunction schematically (omitting transverse coordinates and color indices)  has the form
\beq
|\psi\rangle\,=\, (1\,-\,g_s^2\,\kappa_0\, JJ)\,|\,no\,soft\, gluons \rangle \,+\,g_s \kappa_1\,J\,|\,one\, soft \,gluon\rangle
\eeq
Here $J$ is the color charge density (valence gluons) which emits the  soft gluons at the next step of the evolution. The probability amplitude for single gluon emission $\kappa_1$ is essentially the
Weizsacker-Williams field. A sharp cutoff in longitudinal momenta is implied in the separation between valence and soft modes in the wavefunction.
The $\kappa_0\, JJ$ term is due to normalization of the wavefunction at the order $g^2_s$, $\kappa_0\sim \kappa_1^2$.  
The JIMWLK Hamiltonian is obtained by computing the expectation value of the $\hat S$-matrix operator (expanded to first order in longitudinal phase space):
\beq
H^{JIMWLK}\,=\,\langle \psi|\, \hat S \,-\,1\,|\psi\rangle
\eeq
The fact that  the real term ($JSJ$) and the virtual term ($JJ$) emerge with the very same kernel $K^{LO}$  in eq. (\ref{LO}) is an important consequence of the
wavefunction normalization. The property that $H^{JIMWLK}$ vanishes if we set $S(z)=1$ and $J_L=J_R$  reflects the fact that if none of the particles in the wave function scatter, the scattering matrix does not evolve with energy. This fundamental property must be preserved also at NLO.

To compute the NLO Hamiltonian,  the wavefunction has to be computed to order $g_s^3$ and normalized to order $g_s^4$ \cite{LM}:
 \begin{eqnarray}
|\psi\rangle&=& (1\,-\,g_s^2\,\kappa_0\, JJ\,-\,g_s^4(\delta_1\,JJ\,+\,\delta_2\,JJJ\,+\,\delta_3\,JJJJ)\,|\,no\,soft\, gluons \rangle \,+
\nonumber \\
&&+\,(\,g_s \kappa_1\,J\,+\,g_s^3\epsilon_1\, J\,+\, g_s^3\,\epsilon_2\,J \,J)\,|\,one\, soft \,gluon\rangle
\,+\, g_s^2 (\epsilon_3\, J\,\,+\,\epsilon_4\, JJ)\,|\,two\, soft \,gluons\rangle\,+\,g_s^2\,\epsilon_5\,J\,|\,q\, \bar q\rangle
\end{eqnarray}
With this structure at hand (subtracting second iteration of $H^{LO}$, which is suppressed by an extra power of the longitudinal phase space)
we can write down the most general form of the NLO Hamiltonian, which preserves the $SU_L(N_c)\times SU_R(N_c)$, the signature and the charge conjugation 
symmetries \cite{reggeon}:
\begin{eqnarray}\label{NLO}
&&H^{NLO\ JIMWLK}= \int_{x,y,z}\, K_{JSJ}(x,y;z)\, \left[ J^a_L(x)\,J^a_L(y)\,+\,J_R^a(x)\,J_R^a(y)\,-\,2\,J_L^a(x)\,S_A^{ab}(z)\,J^b_R(y)\right] \ + \nonumber \\
 &&+\int_{x\,y\, z\,z^\prime}\, K_{JSSJ}(x,y;z,z^\prime)\left[f^{abc}\,f^{def}\,J_L^a(x) \,S^{be}_A(z)\,S^{cf}_A(z^\prime)\,J_R^d(y)\,-\, N_c\,J_L^a(x)\,S^{ab}_A(z)\,J^b_R(y)\right] \,
 +\nonumber \\
 &&+\int_{x,y, z,z^\prime} K_{q\bar q}(x,y;z,z^\prime)\left[2\,J_L^a(x) \,tr[S^\dagger(z)\, T^a\,S (z^\prime)T^b]\,J_R^b(y)\,
 -\, J_L^a(x)\,S^{ab}_A(z)\,J^b_R(y)\right]+\nonumber \\
 &&+\int_{w,x,y, z,z^\prime}K_{JJSSJ}(w;x,y;z,z^\prime)f^{acb}\,\Big[J_L^d(x)\, J_L^e(y)\, S^{dc}_A(z)\,S^{eb}_A(z^\prime)\,J_R^a(w)\,-
 \,J_L^a(w)\,S^{cd}_A(z)\,S^{be}_A (z^\prime)\,J_R^d(x)\,J_R^e(y)\,+\nonumber \\
 &&+\frac{1}{3}[\,J_L^c(x) \,J_L^b(y) \,J_L^a(w)\,-\, 
J_R^c(x) \,J_R^b(y)\,J_R^a(w)]\,\Big] \,
 +\nonumber \\
 &&+\int_{w,x,y, z}\, K_{JJSJ}(w;x,y;z)\,f^{bde}\,\Big[  J_L^d(x) \,J_L^e(y) \,S^{ba}_A(z)\,J_R^a(w)\,-\, 
J_L^a(w)\,S^{ab}_A(z)\,J_R^d(x)\,J_R^e(y) \, +\, \nonumber \\
 &&+\frac{1}{3}[\,J_L^d(x) \,J_L^e(y) \,J_L^b(w)\,-\, 
J_R^d(x)\,J_R^e(y)\,J_R^b(w)]
\Big] 
\end{eqnarray}
All $J$s in  (\ref{NLO}) are assumed not to act on $S$ in the Hamiltonian.  
No other color structures appear in the light cone wave function calculation. The discrete symmetries require the kernels $K_{JSSJ}$ and $K_{q\bar q}$
to be symmetric under the interchanges $z\leftrightarrow z^\prime$ or $x\leftrightarrow y$, while  $K_{JJSSJ}$ to be antisymmetric under simultaneous 
 interchange $z\leftrightarrow z^\prime$ and  $x\leftrightarrow y$. We also note that to determine the coefficient of the three $J$ virtual term we need to use the tree level conformal invariance of QCD \cite{complete}.
Our aim is to determine these kernels by comparing the general structure of eq.(\ref{NLO}) to the results of \cite{BC} and \cite{Grab}.

Ref. \cite{BC}  has computed the evolution of a quark-antiquark dipole ${\cal U}=tr[S(u)\,S^\dagger(v)]/N_c$.  In the JIMWLK formalism, this
evolution is generated by acting with $H^{NLO\,JIMWLK}$ on $ {\cal U}$ according to eq.(\ref{1}). The action is defined through the action of the rotation generators $J_L$ 
and $J_R$ (\ref{LR}) and is a purely algebraic operation. The five kernels contribute to the evolution of the dipole and each contribution can be identified in eq. (5) of ref. \cite{BC}. 
It may be possible to recover all  five kernels solely from the evolution of the dipole given in \cite{BC}. It is however more straightforward to supplement this by the results of \cite{Grab}, which  provides an additional piece of NLO calculation. Ref. \cite{Grab} calculated part of the evolution of the three quark singlet amplitude 
$B=\epsilon_{ijk}\epsilon_{lmn}\,S^{im}(u)S^{jl}(v)S^{kn}(w)$ for $N_c=3$, that involves diagrams which couple all three Wilson lines.  In our Hamiltonian these contributions are generated by terms originating from $K_{JJSJ}$ and $K_{JJSSJ}$ where all factors of $J$ act on different Wilson lines. These two kernels 
contribute both to the evolution of the dipole and the three quark singlet. We have determined these kernels using the results of \cite{Grab} and have checked that the results of \cite{BC} are reproduced correctly with our Hamiltonian. This constitutes a non-trivial crosscheck on our calculation, as well as that of refs.\cite{BC} and \cite{Grab}.

We now quote the resulting expressions for the kernels: 
\begin{eqnarray}
K_{JSJ}(x,y;z) =-\frac{\alpha_s^2}{16 \pi^3}
\frac{(x-y)^2}{X^2 Y^2}\Big[b\ln(x-y)^2\mu^2
-b\frac{X^2-Y^2}{ (x-y)^2}\ln\frac{X^2}{Y^2}+
(\frac{67}{9}-\frac{\pi^2}{ 3})N_c-\frac{10}{ 9}n_f\Big]\,-\, \frac{N_c}{2}\ \int_{z^\prime}\, \tilde K(x,y,z,z^\prime)\nonumber \\
\end{eqnarray}
Here $\mu$ is the normalization point in the $\overline{MS}$ scheme and
$b=\frac{11}{3}N_c-\frac{2}{3}n_f$ is the first coefficient of the $\beta$-function.
\begin{eqnarray}
&&K_{JSSJ}(x,y;z,z^\prime) \ =\ \frac{\alpha_s^2}{16\,\pi^4}
\Bigg[\,-\,\frac{4}{ (z-z^\prime)^4}\,+\,
\Big\{2\frac{X^2{Y'}^2+{X'}^2Y^2-4(x-y)^2(z-z')^2}{ (z-z')^4[X^2{Y'}^2-{X'}^2Y^2]}\nonumber\\ 
&&+
~\frac{(x-y)^4}{ X^2{Y'}^2-{X'}^2Y^2}\Big[
\frac{1}{X^2{Y'}^2}+\frac{1}{ Y^2{X'}^2}\Big]
+\frac{(x-y)^2}{(z-z')^2}\Big[\frac{1}{X^2{Y'}^2}-\frac{1}{ {X'}^2Y^2}\Big]\Big\}
\ln\frac{X^2{Y'}^2}{ {X'}^2Y^2}\Bigg]\,+\,\tilde K(x,y,z,z^\prime)
\end{eqnarray}
\beq
\tilde K(x,y,z,z^\prime)\,=\frac{i}{2}\,\left[K_{JJSSJ}(x;x,y;z,z^\prime)-K_{JJSSJ}(y;x,y;z,z^\prime)-K_{JJSSJ}(x;y,x;z,z^\prime)+K_{JJSSJ}(y;y,x;z,z^\prime)\right]
\eeq
\begin{eqnarray}
K_{q\bar q}(x,y;z,z^\prime)\, =\,-\,\frac{\alpha_s^2\,n_f}{ 8\,\pi^4}
\Big\{
\frac{{X'}^2Y^2+{Y'}^2X^2-(x-y)^2(z-z')^2}{ (z-z')^4(X^2{Y'}^2-{X'}^2Y^2)}
\ln\frac{X^2{Y'}^2}{ {X'}^2Y^2}\,-\,\frac{2}{(z-z^\prime)^4}\Big\}
\end{eqnarray}
\begin{eqnarray}
K_{JJSJ}(w;x,y;z)\,=\,-\,i\,\frac{\alpha_s^2}{ 4\, \pi^3 }\,\Big[ \frac{X\cdot W}{ X^2\,W^2}\,-\, \frac{Y\cdot W}{ Y^2\,W^2}    \Big] \ln\frac{Y^2}{ (x-y)^2}\,\ln\frac{X^2}{ (x-y)^2}
\end{eqnarray}
\begin{eqnarray}
K_{JJSSJ}(w;x,y;z,z^\prime)=\,-\,i\,
\frac{\alpha_s^2}{ 2\,\pi^4}
\left(\frac{X_iY^\prime_j}{ X^2Y^{\prime 2}}
\right)\Big(\frac{\delta_{ij}}{2 (z-z^\prime)^2}+\frac{(z^\prime-z)_i W^\prime_j}{ (z^\prime-z)^2 W^{\prime 2}}+
\frac{(z-z^\prime)_j W_i}{ (z-z^\prime)^2 W^{ 2}}-\frac{W_i W^\prime_j}{ W^2 W^{\prime 2}}
\Big)\ln\frac{W^2}{ {W'}^2}
\end{eqnarray}
We note, that as long as one is interested in evolution of color singlet amplitudes, all the kernels are defined only modulo terms that do not depend on (at least) one of the coordinates carried by one of the charge density operators $J$. The integral of $J$ then annihilates any color singlet state and such additional terms do not contribute to the evolution.  The terms proportional to $1/ (z-z^\prime)^4$ and independent of $X$ and $Y$ in $K_{JSSJ}$ and $K_{q\bar q}$  are  such terms. We assigned them to the kernels in this form, so that the kernels vanish at $x=y$ analogously to the dipole kernel at LO. 
Additional input would be necessary to determine evolution of color nonsinglet states.

The $K_{JJSJ}$ term in the Hamiltonian includes an extension of the LO JIMWLK denoted in  \cite{KL}  as JIMWLK+ and analyzed in \cite{nestor}. It is due to coherent emission of a gluon by two classical sources.  The kernel $K_{JJSJ}$ however contains also additional contributions which cannot be interpreted as corrections to the classical Weizsacker-Williams field. 
The  dependence of the kernel $K_{JJSJ}$ on $W$  is   easy to understand - it is just the Weizsacker-Williams field  of a single gluon in the LO wavefunction.  Similarly, $K_{JJSSJ}$ has the same factorized dependence on $X$ and $Y$ corresponding to Weizsacker-Williams fields of two independently emitted gluons.

With the complete NLO Hamiltonian available it should be a straightforward matter to extend many of the LO results. 
In particular, it now becomes a purely algebraic procedure to write down the full Balitsky's hierarchy \cite{Bal} at NLO (and the BKP equation at 
NLO \cite{NLOBKP}), without the need to perform additional next to leading order calculations.   The essential elements entering this hierarchy  have been computed
in a direct NLO computation and are being reported concurrently with our publication \cite{BClast}.
While defining a triple Pomeron \cite{BW} and other Reggeon vertices \cite{last} at NLO might be ambiguous, for any definition chosen, one should be able to read them off 
directly from the Hamiltonian.  We leave these interesting projects for future study.

{\it Note added.}  
When this work was completed we learned about ref. \cite{Simon} which takes approach similar to ours in $N=4$ super Yang Mills theory.

\section{Appendix: Action of the NLO JIMWLK Hamiltonian on the dipole}

The kernels of the Hamiltonian were deduced by comparing actions of the general Hamiltonian on the dipole and on the three-quark operator.
While the connected part (the one that connects all three lines) of the action on the latter  is almost  trivial and we would not present it here, 
the action on the dipole requires some algebra.  To facilitate comparison with the results of \cite{BC} we present the action 
on the dipole $[u^\dagger v]=tr[S^\dagger(u)S(v)]/{N_c}$
\begin{eqnarray}\label{kaction}
&&\int_{x,y,z}K_{JSJ}(x,y,z)J_L^a(x)S_A^{ab}(z)J^b_R(y)[u^\dagger v]=\int_z K_{JSJ}(u,v,z)\left\{\frac{1}{N_c}[u^\dagger v]-[u^\dagger z][z^\dagger v]\right\}\\
&& \int_{x,y}\, K_{JSJ}(x,y,z)\, \left[ J^a_L(x)\,J^a_L(y)\,+\,J_R^a(x)\,J_R^a(y)\right][u^\dagger v]=-4\int_z K_{JSJ}(u,v,z)\frac{N^2_c-1}{2N_c}[u^\dagger v]\\
&&\int_{w,x,y, z}K_{JJSJ}(w;x,y;z)f^{bde}\Big[  J_L^d(x) J_L^e(y) S^{ba}_A(z)J_R^a(w)-
J_L^a(w)S^{ab}_A(z)J_R^d(x)J_R^e(y) \Big][u^\dagger v]=\nonumber\\\
&&~~~~~~~~~~~~=-iN_c\int_z\Bigg[K_{JJSJ}(v;u,v;z)+K_{3,1}(u;v,u;z)\Bigg]\left\{[u^\dagger z][z^\dagger v]-\frac{1}{N_c}[u^\dagger v]\right\}\\
&&\int_{x,y, z,z^\prime}\, K_{JSSJ}(x,y;z,z^\prime)f^{abc}\,f^{def}\,J_L^a(x) \,S^{be}_A(z)\,S^{cf}_A(z^\prime)\,J_R^d(y)[u^\dagger v]\nonumber \\
&&~~~~~~~~~~~~~~~~~~~~~~~~~~
=-\int_{z,z'}K_{JSSJ}(u,v;z,z^\prime)\left\{[u^\dagger z'][z'^\dagger z][z^\dagger v]-[u^\dagger zz'^\dagger v z^\dagger z']\right\} \\
&&\int_{w,x,y, z,z^\prime}K_{JJSSJ}(w;x,y;z,z^\prime)f^{acb}\Big[J_L^d(x) \,J_L^e(y)\, S^{dc}_A(z)\,S^{eb}_A(z^\prime)\,J_R^a(w)\,-\nonumber \\
&&~~~~~~~~~~~~~~~~~~~~~~~~~~~~~~~~~~~~~~~~~~~ -  \,J_L^a(w)\,S^{cd}_A(z)\,S^{be}_A (z^\prime)\,J_R^d(x)\,J_R^e(y)\,\Big][u^\dagger v]=\nonumber\\
 &&~~~~~~~=\frac{i}{2}\int_{z,z^\prime}\Bigg\{\Big[K_{JJSSJ}(v;v,u;z,z')+K_{JJSSJ}(u;u,v;z,z')-K_{JJSSJ}(u;v,u;z,z')-\nonumber \\
 && ~~~~~~~-K_{JJSSJ}(v;u,v;z,z')\Big]  \left\{[u^\dagger z'][z'^\dagger z][z^\dagger v]-[u^\dagger zz'^\dagger v z^\dagger z']\right\}\ +\ 
\Big[K_{JJSSJ}(v;v,u;z,z')-\nonumber \\
&&~~~~~~~-K_{JJSSJ}(u;u,v;z,z')-K_{JJSSJ}(u;v,u;z,z')+K_{JJSSJ}(v;u,v;z,z')-K_{JJSSJ}(v;u,u;z,z')+\nonumber\\
 &&~~~~~~~~~~~~~~~~~~~~~~~~~~~~~~~~~~~~~~~~~+
K_{JJSSJ}(u;v,v;z,z')\Big]\ [u^\dagger z'][z'^\dagger z][z^\dagger v]\Bigg\}\\
 &&\int_{w,x,y} K_{JJSJ}(w,x,y,z)f^{bde}\left[\,J_L^d(x) \,J_L^e(y) \,J_L^b(w)\,-\, 
J_R^d(x)\,J_R^e(y)\,J_R^b(w)\right][u^\dagger v]=\nonumber\\
&&~~~~~~~~=i\frac{N^2_c-1}{2}\Big\{K_{JJSJ}(v,u,v,z)+K_{JJSJ}(u,v,u,z)-K_{JJSJ}(v,v,u,z)-\nonumber \\
&&~~~~~~~~~~~~~~~~~~~~~~~~~~~~~~~~~~-K_{JJSJ}(u,u,v,z)+K_{JJSJ}(v,u,u,z)+K_{JJSJ}(u,v,v,z)\Big\}[u^\dagger v]
\end{eqnarray}
The results quoted in the text are obtained by comparing the above expressions with the explicit calculation of  \cite{BC}.

\section*{Acknowledgments}
We are most grateful to Ian Balitsky who inspired us to think about this project. We thank S. Caron Huot for interesting correspondence. M.L and Y.M. thank the Physics Department of the University of Connecticut for hospitality.
The research was supported by the DOE grant DE-FG02-92ER40716; the EU FP7 grant PIRG-GA-2009-256313; the  ISRAELI SCIENCE FOUNDATION grant \#87277111;  
and the People Program (Marie Curie Actions) of the European Union's Seventh Framework Program FP7/2007-2013/ under REA
grant agreement \#318921;  and the BSF grant \#2012124.

\end{document}